\documentclass[longauth]{aa}

\usepackage{txfonts}
\usepackage{graphicx}
\usepackage{mathtools}
\usepackage{float}
\usepackage{epstopdf}
\usepackage{amsmath}
\usepackage{afterpage}

\bibpunct{(}{)}{;}{a}{}{,} 

\begin{document}

\title{The association of a J-burst with a solar jet}


\author{
D.~E.~Morosan\inst{\ref{1}} 
 \and 
P.~T.~Gallagher\inst{\ref{1}} 
 \and 
R.~A.~Fallows\inst{\ref{2}} 
 \and 
H.~Reid\inst{\ref{3}} 
 \and 
G.~Mann\inst{\ref{4}} 
 \and 
M.~M.~Bisi\inst{\ref{5}} 
 \and 
J.~Magdaleni\'c\inst{\ref{6}} 
 \and 
H.~O. Rucker\inst{\ref{7}} 
 \and 
B.~Thid\'e\inst{\ref{8}} 
 \and 
C.~Vocks\inst{\ref{4}} 
 \and 
J.~Anderson\inst{\ref{9}} 
 \and 
A.~Asgekar\inst{\ref{2}} \and \inst{\ref{10}} 
 \and 
I.~M.~Avruch\inst{\ref{11}} \and \inst{\ref{12}} 
 \and 
M.~E.~Bell\inst{\ref{13}} 
 \and 
M.~J.~Bentum\inst{\ref{2}} \and \inst{\ref{14}} 
 \and 
P.~Best\inst{\ref{15}} 
 \and 
R.~Blaauw\inst{\ref{2}} 
 \and 
A.~Bonafede\inst{\ref{16}} 
 \and 
F.~Breitling\inst{\ref{4}} 
 \and 
J.~W.~Broderick\inst{\ref{2}} 
 \and 
M.~Br\"uggen\inst{\ref{16}} 
 \and 
L.~Cerrigone\inst{\ref{2}} 
 \and 
B.~Ciardi\inst{\ref{17}} 
 \and 
E.~de Geus\inst{\ref{2}} \and \inst{\ref{18}} 
 \and 
S.~Duscha\inst{\ref{2}} 
 \and 
J.~Eisl\"offel\inst{\ref{19}} 
 \and 
H.~Falcke\inst{\ref{20}} \and \inst{\ref{2}} 
 \and 
M.~A.~Garrett\inst{\ref{21}} \and \inst{\ref{2}} \and \inst{\ref{22}} 
 \and 
J.~M.~Grie\ss{}meier\inst{\ref{23}} \and \inst{\ref{24}} 
 \and 
A.~W.~Gunst\inst{\ref{2}} 
 \and 
M.~Hoeft\inst{\ref{19}} 
 \and 
M.~Iacobelli\inst{\ref{2}} 
 \and 
E.~Juette\inst{\ref{25}} 
 \and 
G.~Kuper\inst{\ref{2}} 
 \and 
R. McFadden\inst{\ref{2}} 
 \and 
D.~McKay-Bukowski\inst{\ref{26}} \and \inst{\ref{27}} 
 \and 
J.~P.~McKean\inst{\ref{2}} \and \inst{\ref{12}} 
 \and 
D.~D.~Mulcahy\inst{\ref{21}} 
 \and 
H.~Munk\inst{\ref{28}} \and \inst{\ref{2}} 
 \and 
A.~Nelles\inst{\ref{29}} 
 \and 
E.~Orru\inst{\ref{2}} 
 \and 
H.~Paas\inst{\ref{30}} 
 \and 
M.~Pandey-Pommier\inst{\ref{31}} 
 \and 
V.~N.~Pandey\inst{\ref{2}} 
 \and 
R.~Pizzo\inst{\ref{2}} 
 \and 
A.~G.~Polatidis\inst{\ref{2}} 
 \and 
W.~Reich\inst{\ref{32}} 
 \and 
D.~J.~Schwarz\inst{\ref{33}} 
 \and 
J.~Sluman\inst{\ref{2}} 
 \and 
O.~Smirnov\inst{\ref{34}} \and \inst{\ref{35}} 
 \and 
M.~Steinmetz\inst{\ref{4}} 
 \and 
M.~Tagger\inst{\ref{23}} 
 \and 
S.~ter Veen\inst{\ref{2}} 
 \and 
S.~Thoudam\inst{\ref{20}} 
 \and 
M.~C.~Toribio\inst{\ref{22}} \and \inst{\ref{2}} 
 \and 
R.~Vermeulen\inst{\ref{2}} 
 \and 
R.~J.~van Weeren\inst{\ref{36}} 
 \and 
O.~Wucknitz\inst{\ref{32}} 
 \and 
P.~Zarka\inst{\ref{37}}
}

\institute{
School of Physics, Trinity College Dublin, Dublin 2, Ireland \label{1}
\and
ASTRON, Netherlands Institute for Radio Astronomy, Postbus 2, 7990 AA, Dwingeloo, The Netherlands \label{2}
\and
School of Physics and Astronomy, SUPA, University of Glasgow, Glasgow G12 8QQ, United Kingdom \label{3}
\and
Leibniz-Institut f\"{u}r Astrophysik Potsdam (AIP), An der Sternwarte 16, 14482 Potsdam, Germany \label{4}
\and
RAL Space, Science and Technology Facilities Council, Rutherford Appleton Laboratory, Harwell Campus, Oxfordshire, OX11 OQX, United Kingdom \label{5}
\and
Solar-Terrestrial Center of Excellence, SIDC, Royal Observatory of Belgium, Avenue Circulaire 3, B-1180 Brussels, Belgium \label{6}
\and
Commission for Astronomy, Austrian Academy of Sciences, Schmiedlstrasse 6, 8042 Graz, Austria \label{7}
\and
Swedish Institute of Space Physics, Box 537, SE-75121 Uppsala, Sweden \label{8}
\and
Helmholtz-Zentrum Potsdam, DeutschesGeoForschungsZentrum GFZ, Department 1: Geodesy and Remote Sensing, Telegrafenberg, A17, 14473 Potsdam, Germany \label{9}
\and
Shell Technology Center, Bangalore, India \label{10}
\and
SRON Netherlands Insitute for Space Research, PO Box 800, 9700 AV Groningen, The Netherlands \label{11}
\and
Kapteyn Astronomical Institute, PO Box 800, 9700 AV Groningen, The Netherlands \label{12}
\and
CSIRO Astronomy and Space Science, 26 Dick Perry Avenue, Kensington, WA 6151, Australia  \label{13}
\and
University of Twente, The Netherlands \label{14}
\and
Institute for Astronomy, University of Edinburgh, Royal Observatory of Edinburgh, Blackford Hill, Edinburgh EH9 3HJ, UK \label{15}
\and
University of Hamburg, Gojenbergsweg 112, 21029 Hamburg, Germany \label{16}
\and
Max Planck Institute for Astrophysics, Karl Schwarzschild Str. 1, 85741 Garching, Germany \label{17}
\and
SmarterVision BV, Oostersingel 5, 9401 JX Assen \label{18}
\and
Th\"{u}ringer Landessternwarte, Sternwarte 5, D-07778 Tautenburg, Germany \label{19}
\and
Department of Astrophysics/IMAPP, Radboud University Nijmegen, P.O. Box 9010, 6500 GL Nijmegen, The Netherlands \label{20}
\and
Jodrell Bank Center for Astrophysics, School of Physics and Astronomy, The University of Manchester, Manchester M13 9PL,UK \label{21}
\and
Leiden Observatory, Leiden University, PO Box 9513, 2300 RA Leiden, The Netherlands \label{22}
\and
LPC2E - Universite d'Orleans/CNRS \label{23}
\and
Station de Radioastronomie de Nancay, Observatoire de Paris - CNRS/INSU, USR 704 - Univ. Orleans, OSUC , route de Souesmes, 18330 Nancay, France \label{24}
\and
Astronomisches Institut der Ruhr-Universit\"{a}t Bochum, Universitaetsstrasse 150, 44780 Bochum, Germany \label{25}
\and
Sodankyl\"{a} Geophysical Observatory, University of Oulu, T\"{a}htel\"{a}ntie 62, 99600 Sodankyl\"{a}, Finland \label{26}
\and
STFC Rutherford Appleton Laboratory,  Harwell Science and Innovation Campus,  Didcot  OX11 0QX, UK \label{27}
\and
Radboud University Radio Lab, Nijmegen, P.O. Box 9010, 6500 GL Nijmegen, The Netherlands \label{28}
\and
Department of Physics and Astronomy, University of California Irvine, Irvine, CA 92697, USA \label{29}
\and
Center for Information Technology (CIT), University of Groningen, The Netherlands \label{30}
\and
Centre de Recherche Astrophysique de Lyon, Observatoire de Lyon, 9 av Charles Andr\'{e}, 69561 Saint Genis Laval Cedex, France \label{31}
\and
Max-Planck-Institut f\"{u}r Radioastronomie, Auf dem H\"{u}gel 69, 53121 Bonn, Germany \label{32}
\and
Fakult\"{a}t f\"{u}r Physik, Universit\"{a}t Bielefeld, Postfach 100131, D-33501, Bielefeld, Germany \label{33}
\and
Department of Physics and Elelctronics, Rhodes University, PO Box 94, Grahamstown 6140, South Africa \label{34}
\and
SKA South Africa, 3rd Floor, The Park, Park Road, Pinelands, 7405, South Africa \label{35}
\and
Harvard-Smithsonian Center for Astrophysics, 60 Garden Street, Cambridge, MA 02138, USA \label{36}
\and
LESIA \& USN, Observatoire de Paris, CNRS, PSL/SU/UPMC/UPD/SPC, Place J. Janssen, 92195 Meudon, France \label{37}
}

\date{ Received /
		Accepted }

\abstract{The Sun is an active star that produces large-scale energetic events such as solar flares and coronal mass ejections and numerous smaller-scale events such as solar jets. These events are often associated with accelerated particles that can cause emission at radio wavelengths. The reconfiguration of the solar magnetic field in the corona is believed to be the cause of the majority of solar energetic events and accelerated particles. }
{Here, we investigate a bright J-burst that was associated with a solar jet and the possible emission mechanism causing these two phenomena. }
{We used data from the Solar Dynamics Observatory (SDO) to observe a solar jet, and radio data from the Low Frequency Array (LOFAR) and the Nan{\c c}ay Radioheliograph (NRH) to observe a J-burst over a broad frequency range (33--173~MHz) on 9 July 2013 at $\sim$11:06 UT. }
{The J-burst showed fundamental and harmonic components and it was associated with a solar jet observed at extreme ultraviolet wavelengths with SDO. The solar jet occurred at a time and location coincident with the radio burst, in the northern hemisphere, and not inside a group of complex active regions in the southern hemisphere. The jet occurred in the negative polarity region of an area of bipolar plage. Newly emerged positive flux in this region appeared to be the trigger of the jet.  }
{Magnetic reconnection between the overlying coronal field lines and the newly emerged positive field lines is most likely the cause of the solar jet. Radio imaging provides a clear association between the jet and the J-burst which shows the path of the accelerated electrons. These electrons travelled from a region in the vicinity of the solar jet along closed magnetic field lines up to the top of a closed magnetic loop at a height of $\sim$360~Mm. Such small-scale complex eruptive events arising due to magnetic reconnection could facilitate accelerated electrons continuously to produce the large numbers of Type III bursts observed at low frequencies, in a similar way to the J-burst analysed here. }


\maketitle

\section{Introduction}

{The reconfiguration of the solar magnetic field in the corona is believed to be the cause of the majority of solar energetic events such as solar jets, flares and coronal mass ejections (CMEs). The energy released during these events contributes to the acceleration of particles that travel along magnetic field lines away from the energy release site \citep{vi02}.  }

{Flares and solar jets are associated with the production of X-ray emission and Type III radio bursts \citep{ku94,ra96, kr11} and the injection of near-relativistic electrons into interplanetary space \citep{li85}. Solar jets are smaller-scale eruptions of material compared to flares and CMEs, that have been well-observed at X-ray wavelengths \citep{kr11}. Accelerated electrons that are responsible for X-ray and radio emission can travel up towards the high corona and down towards the chromosphere, away from the energy release site. Downward travelling electrons would lose their energy due to collisions producing hard X-ray (HXR) footpoints \citep{kr11}. Upward travelling electrons can produce Type III radio bursts that can be identified as rapidly varying bursts of radiation in dynamic spectra with durations of up to a few seconds \citep{wild50}. Type III bursts are considered to be the radio signature of electron beams travelling through the corona and into interplanetary space along open and quasi-open magnetic field lines \citep{li74}. It is commonly believed that accelerated electrons outpace the slower background electrons producing a bump-on-tail instability in their velocity distribution. This generates Langmuir (plasma) waves \citep{ro93} which are then converted into radio waves at the local plasma frequency ($f_p$) and its harmonic \citep[$2f_p$;][]{ba98}. }

{Accelerated electron beams in the solar corona are generally considered to be generated during magnetic reconnection or accelerated at shock wave fronts \citep{du00}. \citet{kr11} studied the interchange reconnection scenario where a magnetic flux emerges in the corona near open magnetic field lines with which it reconnects. In this scenario, three HXR footpoints were predicted to occur which were also found in some of the events studied. \citet{gl12} showed that Type III radio bursts also occur as a result of interchange reconnection since the accelerated electrons have access to open magnetic field lines. Other studies by \citet{in11} and \citet{ch13} also observed Type III radio bursts and HXR emission associated with EUV jets. However, not all bursts escape along open magnetic field lines; a type U-burst has also been associated with a plasma jet \citep{au94}. J-bursts or U-bursts are variants of Type III radio bursts and represent the radio signature of electron beams travelling along closed magnetic loops \citep{ma58, st71}. }

\begin{figure*}[!ht] 
\centering
\includegraphics[angle = 0, width = 312px, trim = 15px 20px 25px 10px ]{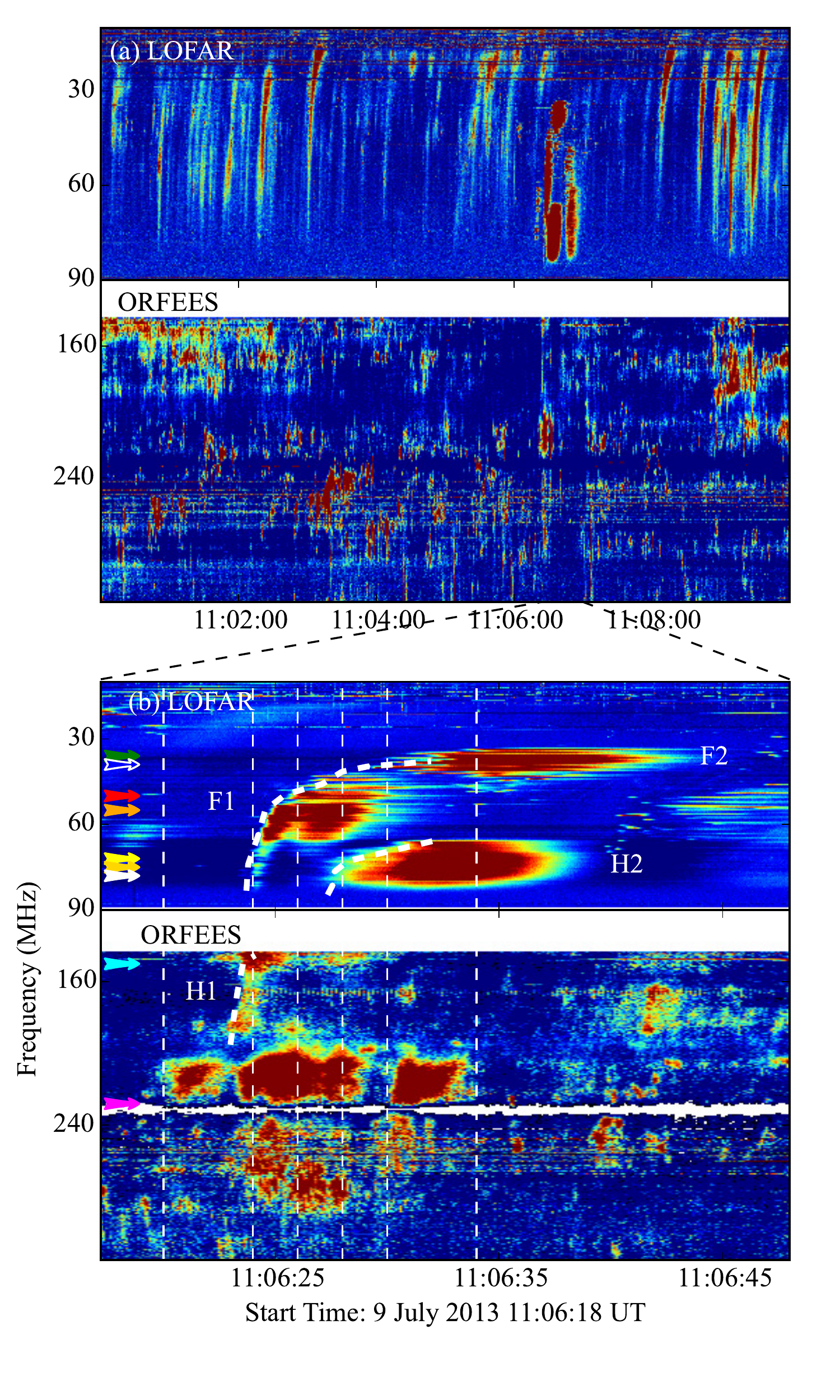}
\caption{  (a) Composite dynamic spectrum at frequencies of 10--240~MHz (top panel) showing multiple Type III radio bursts observed by LOFAR (top) and a Type I noise storm observed by ORFEES (bottom) on 9 July 2013 starting from 11:00 to 11:10~UT. (b) Zoom in of the dynamic spectrum in (a) focusing on the bright J-burst observed by LOFAR (top) and ORFEES (bottom). The J-burst is denoted by white dashed lines and its components are labelled F1, F2, H1 and, H2. The vertical lines and colour-coded arrows denote the time and frequency, respectively, of the bursts shown in Figure 2. \label{fig1}}
\end{figure*} 

\begin{figure*}[t]
\centering
\includegraphics[angle = 0, width = 440px, trim = 0px 0px 0px 0px ]{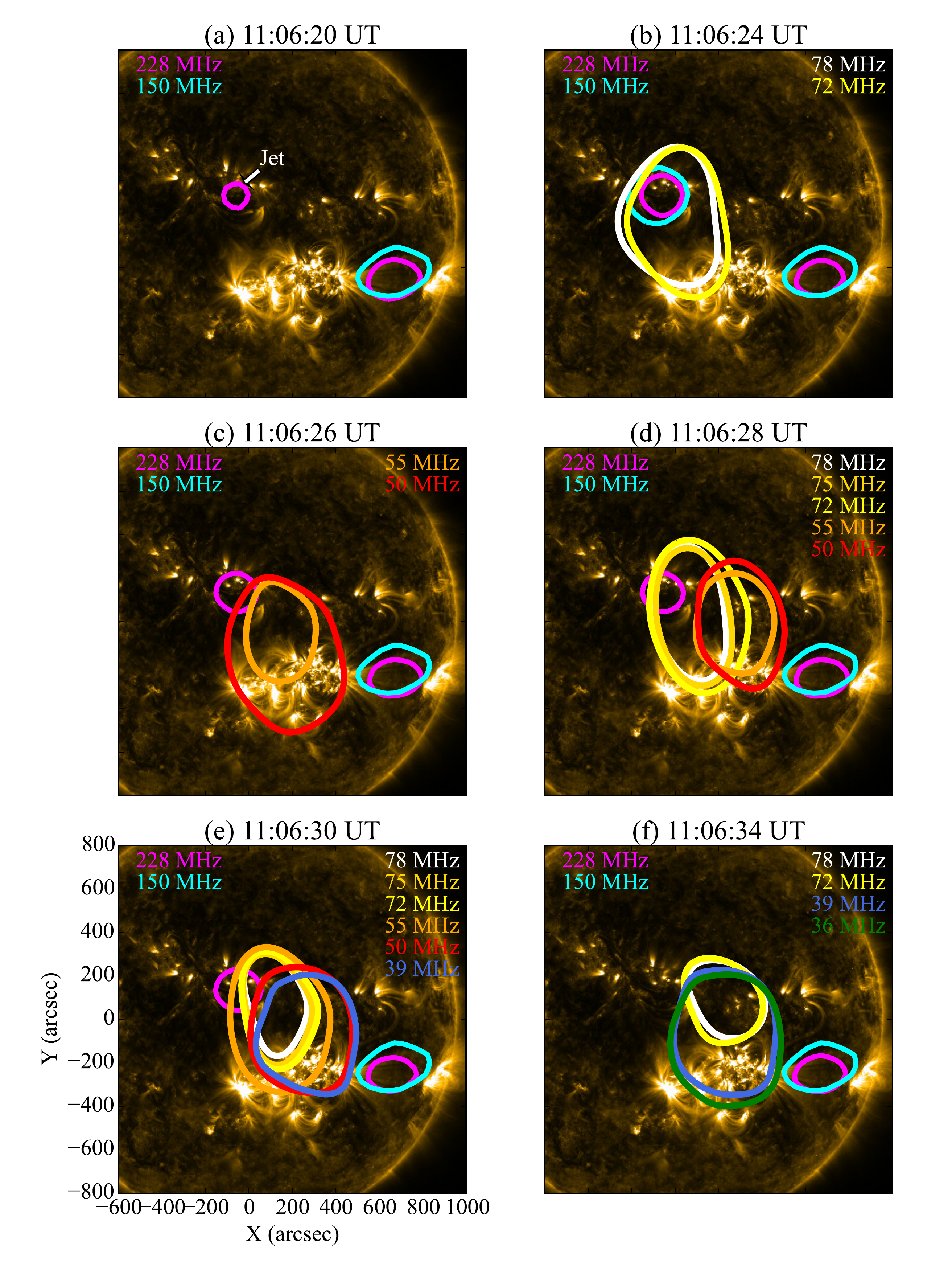}
\caption{ AIA 171~\AA~ image of the corona overlaid with NRH and LOFAR sources in order of appearance at (a) 11:06:20 UT, (b) 11:06:24 UT, (c) 11:06:26 UT, (d) 11:06:28 UT, (e) 11:06:30 UT and (f) 11:06:34 UT. These times are denoted by dashed lines in Figure 1 for comparison with the dynamic spectrum. F1 and H1 are overlaid in panel (b) while H2 and F2 are overlaid in panels (e) and (f). \label{fig2}}
\end{figure*} 

{Emerging flux reconnection models have been proposed as the mechanism responsible for the initiation of solar flares \citep{he77, he78}. These models were later expanded by \citet{sh92} showing that the magnetic reconnection between an emerging flux and the overlying coronal magnetic field leads to the ejection of cool and hot jets, magnetohydrodynamic (MHD) shocks, compact flares, and X-ray bright points. Numerical simulations by \citet{sh92} showed that, as a result of reconnection, hot plasma jets form due to enhanced gas pressure which can be observed at X-ray and extreme ultraviolet (EUV) wavelengths. A cool plasma jet can also be ejected due to cool, dense chromospheric plasma confined in magnetic islands formed in the current sheet due to flux emergence \citep{yo96}. Cool jets have been observed as H$\alpha$ surges often accompanying the X-ray jets \citep{shi96}. In more recent observations, \citet{st15} showed that small-scale filament eruptions are the drivers of X-ray jets in coronal holes as opposed to the emerging flux model. }

{Despite a number of studies and simulations of solar jets associated with radio emission, there are still unanswered questions about their exact mechanism and the path followed by accelerated electrons in the solar corona. In this paper, we aim to investigate the mechanism responsible for the production of a J-burst associated with a solar jet observed on 9 July 2013. In Section 2 we give an overview of the observations. In Section 3 we present the results of this study which are further discussed in Section 4. }

\section{Observations}

\begin{figure*}[!ht] 
\centering
\includegraphics[angle = 90, width = 330px, trim = 0px 20px 0px 10px ]{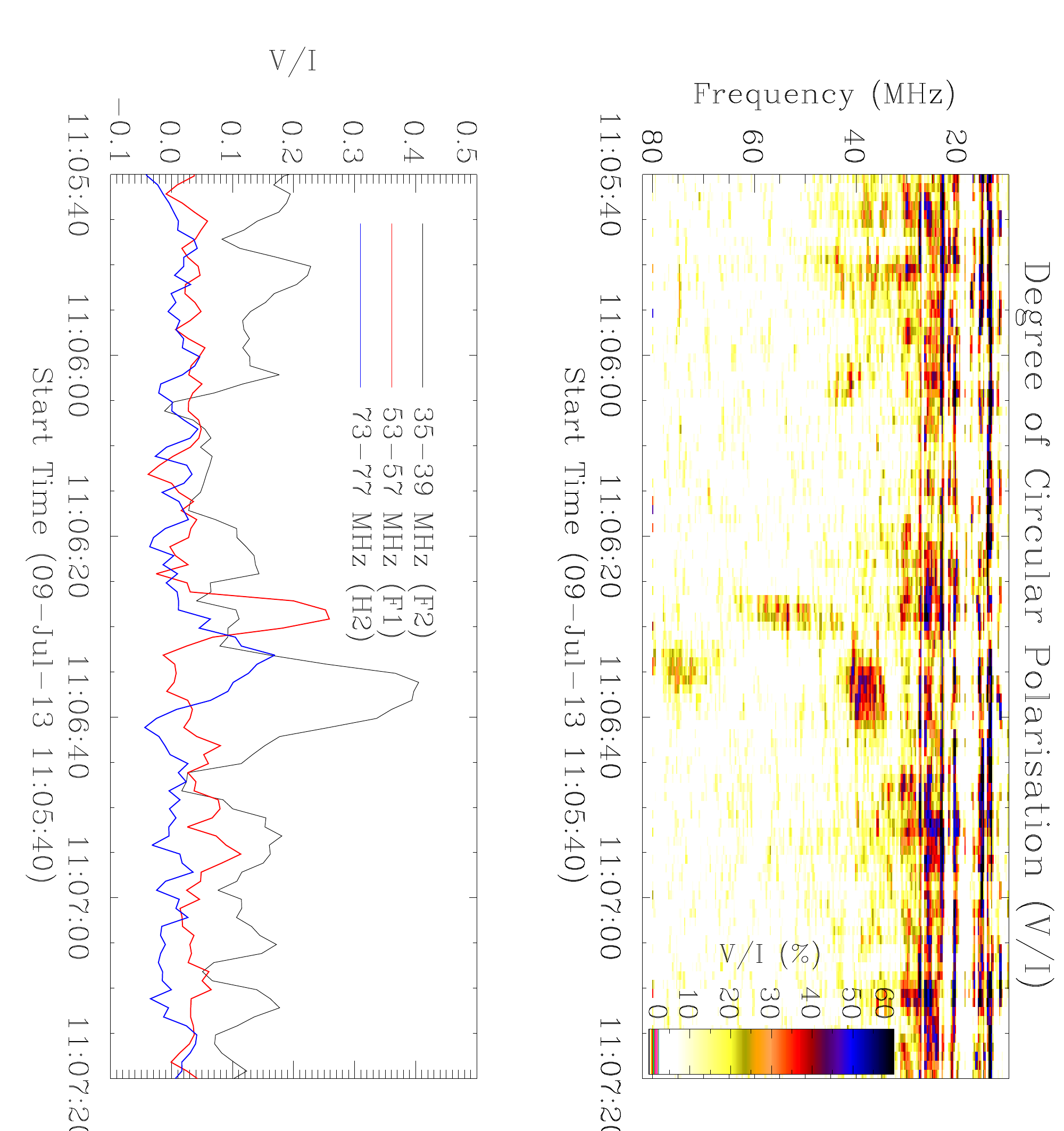}
\caption{  Degree of circular polarisation of the Type III-like burst components (F1, F2 and H2) as observed by the Nan{\c c}ay Decamteric Array \citep[NDA;][]{bo80}. \textit{Top panel:} NDA dynamic spectrum of the degree of circular polarisation of the Type III-like burst. \textit{Bottom panel:} Frequency slices from the top panel averaged over 20 frequency bins to produce the average degree of polarisation over frequency ranges that sample F2, F1 and H2, respectively.  \label{fig1}}
\end{figure*}

\begin{figure*}[ht]
\centering
\includegraphics[angle = 0, width = 420px, trim = 0px 20px 0px 20px ]{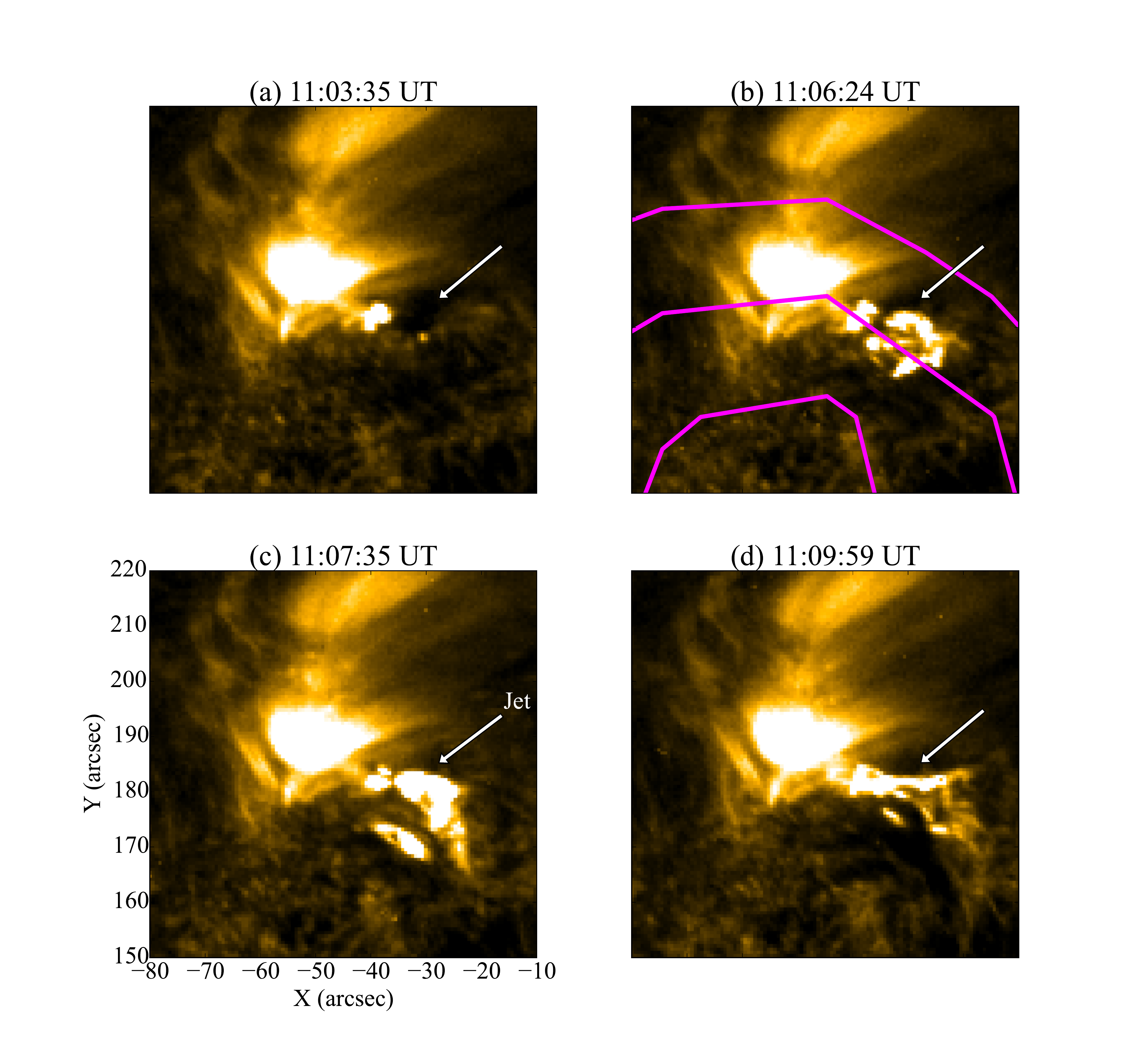}
\caption{Evolution of the solar jet observed on 9 July 2013 as seen at four separate times in the AIA 171~\AA~ channel. The white arrow points to the location of the solar jet. The jet starts to erupt at $\sim$11:06:23 UT which is the time when the plasma motion is directed southwards of the source region. The solar jet can also be seen in the movie accompanying this paper. The purple contours in (b) represent the 70\%, 80\% and 90\% contour levels of the 228~MHz radio source at 11:06:24 UT.\label{fig3}}
\end{figure*}

{On 9 July 2013 at 11:06:25 UT, an unusual, bright Type III-like radio burst was observed by the Low Frequency Array \citep[LOFAR;][]{lofar13} at frequencies of 33--80~MHz as seen in Figure 1a. The burst is similar in appearance and characteristics to other Type III bursts, however, it is unusual as it is composed of a few components  and the emisison suddenly stops at a frequency of 33~MHz as seen in Figure 1b. The burst is therefore named a J-burst throughout this analysis. The J-burst was superimposed on a multitude of fainter Type III radio bursts as seen in Figure 1a. }

{LOFAR is a new-generation radio interferometric array which consists of thousands of dipole antennas distributed in 24 core stations and 14 remote stations throughout the Netherlands and 13 international stations across Europe. In this study we have used observations from the Low Band Antennas (LBAs) of the LOFAR core that operate at frequencies of 10-90~MHz. Using one of LOFAR's beam-formed modes, we produced 170 simultaneous tied-array beams to image the Sun at LBA frequencies \citep{st11}. Each tied-array beam had a FWHM ranging between 7\arcmin~ at 90 MHz to 21\arcmin~ at 30 MHz and it recorded a high-time and frequency resolution dynamic spectrum ($\sim$10~ms; 12.2 kHz). The tied-array beams covered a field of view of 1.3\degr, centred on the Sun. Tied-array imaging was used to image the J-burst observed by LOFAR at several frequencies as seen in Figure 2 \citep[for more details concerning the technique see][]{mo14, mo15}. }

{The J-burst was also observed at higher frequencies between 140--240~MHz in the ORFEES (Observation Radio Frequence pour l'Etude des Eruptions Solaires) dynamic spectra (Figure 1c) where it also shows multiple components. In addition it was also identified in images from the the Nan{\c c}ay Radioheliograph \citep[NRH;][]{ke97} at frequencies of 150 (Figure 2), 173, 228 (Figure 2), 270 and 298~MHz. ORFEES is a new radio-spectrograph located in Nan{\c c}ay, France, observing between 140 and 1000 MHz and NRH is a solar dedicated interferometric array of dishes, also located in Nan{\c c}ay, capable of imaging the Sun at several frequencies ranging from 150 to 445~MHz. Figure 2 shows the NRH 50\% level contours at frequencies of 150~MHz and 228~MHz. At these frequencies, the J-burst occurred simultaneously with a Type I noise storm (which can be seen by the fine structure noise-like features in Figure 1b and the NRH radio burst contours to the right in the panels of Figure 2). The Type I noise storm was associated with the active region NOAA 11788 in the southern hemisphere and it accompanied the active region during its rotation on the visible side of the solar disc. }

{The degree of circular polarisation of the J-burst was analysed with the Nan{\c c}ay Decametric Array \citep[NDA;][]{bo80} as seen in Figure 3 and from NRH images. The NDA is a phased array consisting of 144 antennas. The array is split in two groups of 72 left-handed and 72 right-handed wound antennae to measure the degree of circular polarisation which was used in this study.  }

{A solar jet was observed at EUV wavelengths by the Solar Dynamics Observatory \citep[SDO;][]{pe12} at a time and location coincident to the J-burst. The evolution of the solar jet can be seen in a zoomed-in field of view in Figure 4. The jet was observed in the northern hemisphere of the Sun in an area of bipolar plage by the Atmospheric Imaging Assembly \citep[AIA;][]{le12} onboard SDO. The AIA instrument provides an unprecedented view of the solar corona at multiple wavelengths, imaging the Sun up to 1.1~R$_\sun$. The solar jet was observed in all six AIA coronal filters from 94 to 304~\AA. The eruption of the material commenced at 11:06~UT, which is the time when the plasma was ejected southward of the source region. Afterwards, a clear movement can be seen in the movie accompanying this paper. The brightening of the plasma in the vicinity of the jet started $\sim$2~minutes before the jet onset. The jet lasted for $\sim$8 minutes. Smaller eruptions and brightenings occurred before and after the main event, however, these were not associated with any radio emission.}

\begin{figure*}[!ht] 
\centering
\includegraphics[angle = 0, width = 312px, trim = 15px 30px 25px 20px ]{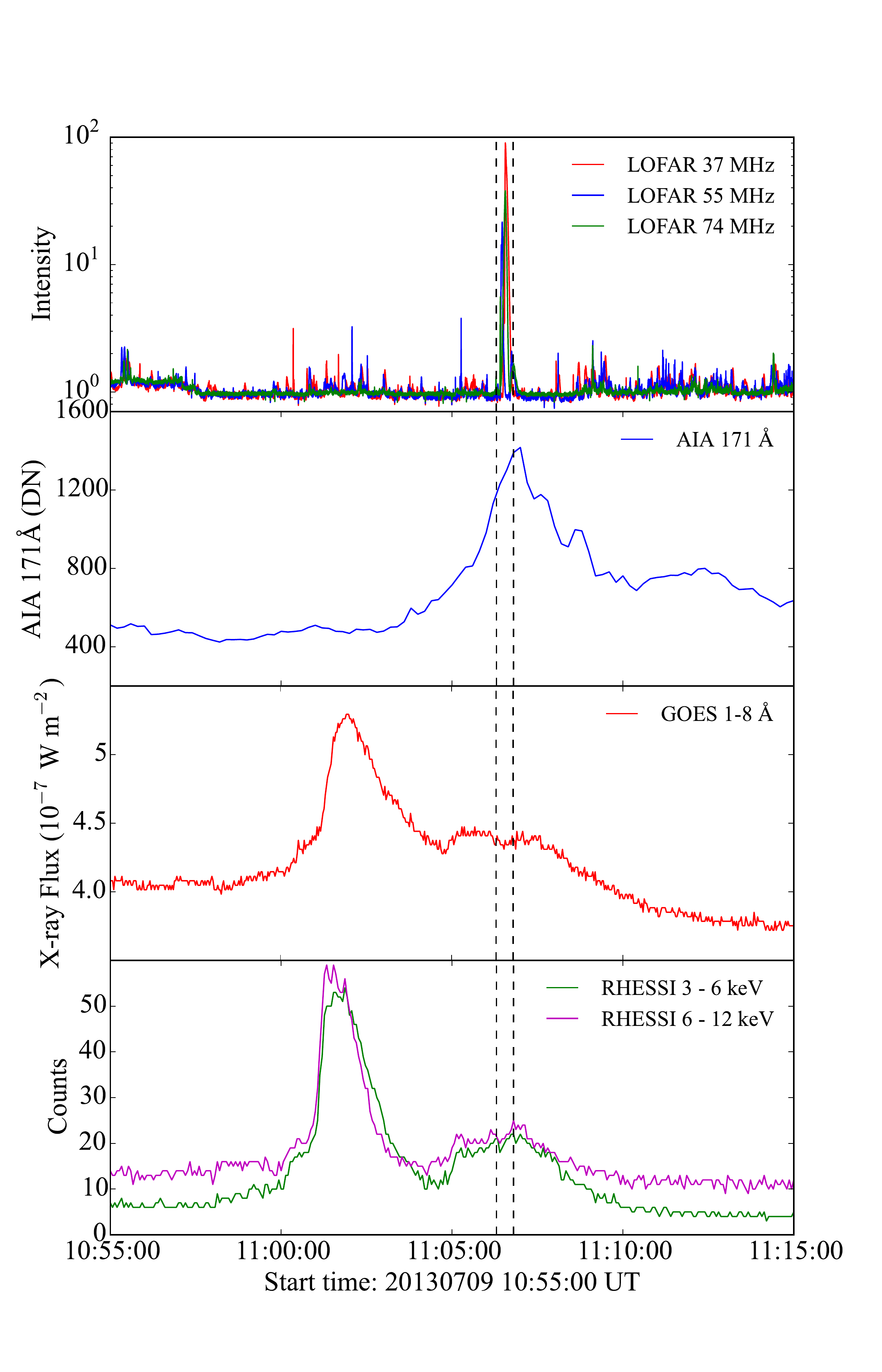}
\caption{  Time series over three frequency channels of J-burst, observed as the highest intensity peaks in the top panel compared to the evolution of the jet in AIA 171~\AA~(second panel), a B-class solar flare in GOES 1--8~\AA~(third panel) and RHESSI 3--6 and 6--12 keV channels (bottom panel). The flare seen in GOES and RHESSI light curves occurs in the southern hemisphere, while the jet and Type III bursts occurred in the northern hemisphere.  The vertical dashed lines denote the time span of the dynamic spectra in Figure 1b.  \label{fig1}}
\end{figure*}

{In order to determine the geometry of the jet, we studied the magnetic geometry of the surrounding area of bipolar plage from which the jet originates. This was done using images from the Helioseismic and Magnetic Imager \citep[HMI;][]{sc12} onboard SDO and potential field source surface models (PFSS) used to study the magnetic field lines configuration in the region \citep{sc03} that use as a boundary the observed heliospheric magnetic field from HMI. HMI is an instrument designed to study the magnetic field at the photosphere, the origin and evolution of sunspots, active regions, and solar activity. }

\begin{figure}[ht]
\centering
\includegraphics[angle = -90, width = 330px, trim = 20px 150px 40px 0px ]{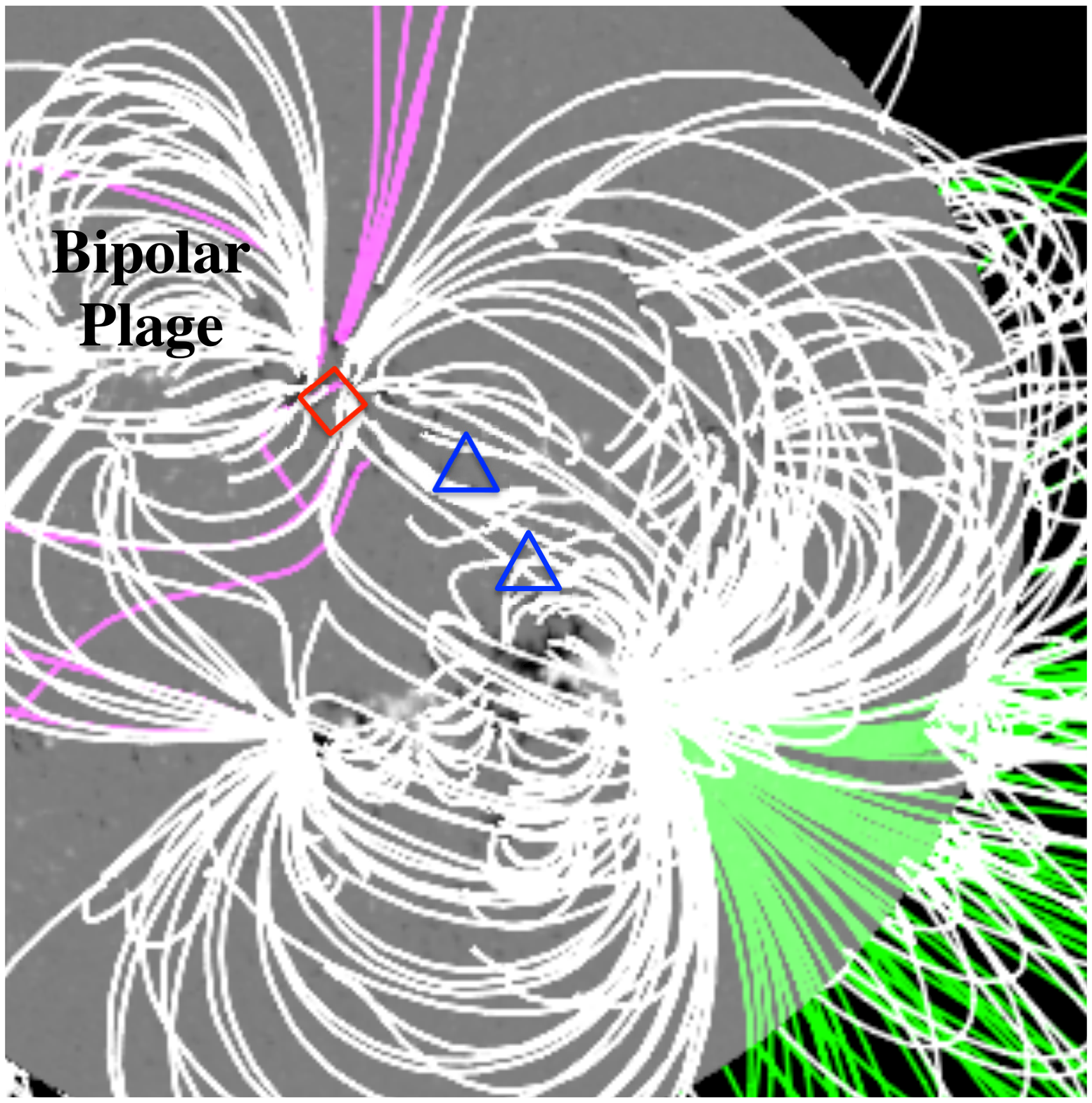}
\caption{Photospheric magnetic field as seen by the HMI magnetogram overlaid with the potential field source surface (PFSS) model of magnetic field lines \citep{sc03}. The position of the NRH 150~MHz (red) and LOFAR 78 and 39~MHz (blue) source centroids at 11:06:30~UT from Figure 2 are overlaid. The solar jet originates in a region below the NRH sources. \label{fig4}}
\end{figure} 

\section{Results}
\subsection{J-burst characteristics}

{The J-burst in Figure 1b was first observed in NRH images and in ORFEES dynamic spectra at 11:06:19~UT at frequencies of 140--298~MHz. At 11:06:24 UT, it was observed with LOFAR LBAs at frequencies of 33--80~MHz. The J-burst is similar in appearance and characteristics to a regular Type III radio burst, however, it is composed of numerous components as seen in Figure 1b. The J-burst is much brighter and more distinctive than the concurrent Type III storm at low frequencies (10--90~MHz) in Figure 1a. }

{The J-burst has a typical drift rate of -9~MHz s$^{-1}$ at frequencies between 45--80~MHz which is comparable with other findings (for example \citealt{ma02} reported a mean Type III drift rate of -11~MHz s$^{-1}$ at a frequency range of 40--70 MHz) and a typical velocity of $\sim$0.2~c. The drift rate and velocity were estimated along the continuous part of the burst (F1). The J-burst suddenly ends at a frequency of 33~MHz unlike all the other Type III bursts observed during the same time period (Figure 1a). The 33~MHz cut-off frequency was also seen in NDA observations. In addition, the burst was absent in the high resolution URAN-2 data \citep{br05} that was observing the Sun at the same time at frequencies between 10--30~MHz. The concurrent weaker Type III bursts in Figure 1a have been observed with URAN-2 and their flux densities were at most 10 s.f.u. (where 1~s.f.u. represents a solar flux unit equal to $10^4$~Jy). All components of the J-burst are much brighter than the concurrent Type III bursts at 10--90~MHz by approximately one to two orders of magnitude. The detailed structure of the J-burst appears very fragmented and multiple components can be identified (for example F1, F2, H1 and, H2, which will be discussed in more detail in the following paragraphs). The correlation among all these components is established by imaging.}

{The J-burst components observed by LOFAR (F1, F2 and H2) at frequencies below 90~MHz were imaged using LOFAR tied-array imaging described in \citet{mo14,mo15}. The higher frequency components ($>$140~MHz) of the J-burst were imaged using the NRH. Figure 2 shows radio contours at six different times to sample the evolution of the J-burst: (a) 11:06:20, (b) 11:06:24, (c) 11:06:26, (d) 11:06:28, (e) 11:06:30 and, (f) 11:06:34. The radio contours are overlaid on top of the AIA 171~\AA ~ image at 11:06:24~UT in (a), (b) and (c) and 11:06:35~UT in (d), (e) and (f). The times in Figure 2 are denoted by the dashed lines in Figure 1 for comparison with the dynamic spectrum and the frequencies of these radio bursts are also denoted by colour-coded arrows in Figure 1. The NRH contours in Figure 2 are plotted at the 50\% contour level while the LOFAR sources are plotted at the 80\% contour level. A higher contour level was plotted for the LOFAR sources as the tied-array beam sizes are quite large ($\sim$8\arcmin~at 78 MHz and $\sim$16\arcmin~at 39 MHz), the beams overlap at low frequencies and the overall source sizes appear large.}

{Radio sources other than the Type I noise storm are first seen in NRH images at higher frequencies (for example 228~MHz) at 11:06:20~UT in Figure 2a. This source, however, does not appear to be part of the J-burst as it does not have the same characteristics in Figure 1b. In addition, this source lasts for $\sim$10~s at frequencies around 228~MHz with no obvious frequency drift, while the J-burst components have a frequency drift rate. All radio sources at frequencies of 228, 270 and 298~MHz (other than the Type I storm) occur in the same spatial location, as the J-burst in Figure 2a, and have a high degree of circular polarisation of up to 50\% based on NRH Stokes V images. The high degree of polarisation and the fact that sources above 228~MHz occur at the start frequency of the J-burst are both characteristics of radio spikes \citep{be82, be96}. Unfortunately the ORFEES dynamic spectra are not sufficient to provide any details of the fine structures of these sources.}

{Four seconds after the appearance of the highly-polarised, high-frequency sources in Figure 2a, the J-burst appears at 11:06:24 UT over a wide bandwidth (Figure 2b). The 150~MHz NRH source in the northern hemisphere that occurs at 11:06:24 UT is unpolarised, indicating that it is not related to the sources above frequencies of 228~MHz. This source is the H1 component of the J-burst based on the dynamic spectrum in Figure 1b. Superimposed on the NRH source, two LOFAR sources can be seen at 11:06:24~UT in Figure 2b at frequencies of 72 and 78~MHz, that represent F1 from Figure 1b. From 11:06:26~UT (Figure 2c), the 150~MHz NRH source disappears in the northern hemisphere and we only see LOFAR bursts at 50 and 55~MHz, which agrees with the ORFEES dynamic spectrum in Figure 1b. At 11:06:28~UT (Figure 2d), the LOFAR sources at 50, 55, 72, 75 and 78~MHz, now, appear at a slightly different spatial position shifted to the right compared to Figure 2b. This indicates that at this time we are sampling another electron beam accelerated in a different direction. At 11:06:30 in Figure 2e, the radio sources at 78, 75, 72, 55, 50 and 39~MHz appear to move in a south-westerly direction from the initial location with decreasing frequency. At 11:06:34~UT, only the LOFAR sources H2 and F2 are present (Figure 2f). }

{The movement and appearance of sources indicates that an initial electron beam was accelerated to produce the burst at 11:06:24 and 11:06:26~UT and then another electron beam escaped along neighbouring magnetic field lines to produce the burst which is offset to the right seen at 11:06:28~UT. These bursts eventually reach a region where these electron beam stops producing radio emission. The magnetic field lines along which these electrons escape are most likely closed since radio imaging shows that the low frequency radio bursts which occur high in the corona are located in between two regions of increased magnetic flux: an area of bipolar plage in the northern hemisphere where the J-burst originates and a complex $\beta\gamma\delta$ active region in the south. }

\begin{figure*}[ht]
\centering
\includegraphics[angle = 0, width = 580px, trim = 60px 70px 0px 70px ]{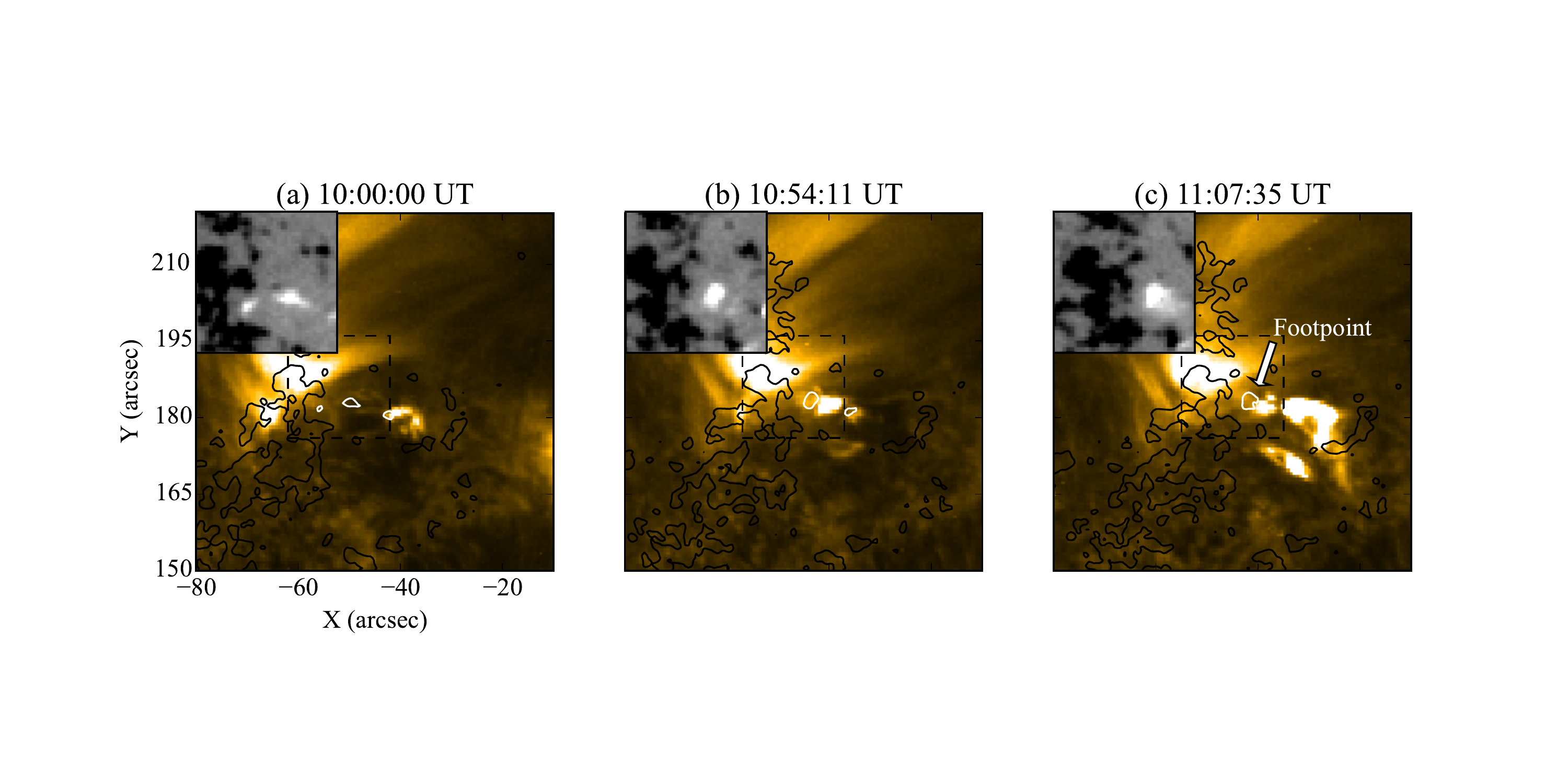}
\caption{Zoomed-in view of the solar jet region where the positive (white) and negative (black) photospheric magnetic field can be seen as contours overlaid on the AIA 171~\AA~images. The contours are drawn at 150~G. The inset on the top left of each panel represents a zoomed-in view of the photospheric magnetic field from HMI in the dashed rectangular box. The positive polarity footpoint responsible for the jet activity is marked by the white arrow in (c). \label{fig4}}
\end{figure*}

{ The-burst components labelled F1 and F2 are patchy and fragmented in the dynamic spectrum (Figure 1b) and also much brighter than the background bursts, which is indicative of fundamental emission. The components labelled H1 and H2 are more diffuse and less bright than F1 and F2, which is indicative of harmonic emission. F1 and H1 occur at approximately the same spatial location at 11:06:24~UT (Figure 2b). F2 and H2 also occur at the same spatial location but to the right of F1 and H1 at 11:06:34~UT (Figure 2f). In addition, the cut-off frequency of F2 (33~MHz) is half the cut-off frequency of H2 (66~MHz), indicating that F2 and H2 are a fundamental-harmonic pair where F2 is emitted at the local plasma frequency and H2 is emitted at twice that frequency. }

{In order to verify that they are indeed a fundamental-harmonic pair it is necessary to look at the polarisation of these bursts. According to \citet{du80}, structureless harmonic components of Type III radio bursts have a low degree of circular polarisation of $<$$6\%$, while fundamental-harmonic pairs are more polarised: the fundamental component is on average 35\% circularly polarised while the harmonic component is on average 11\% polarised. The J-burst observed in this study is also observed with NDA which provides left and right linear polarisations that can be used to measure the degree of circular polarisation (Figure 3). The ratio of Stokes V/I is shown in the top panel of Figure 3 can show the percentage of circular polarisation, while the bottom panel of Figure 3 shows the degree of circular polarisation inside frequency slices that sample F1, F2 and H2. The NDA data shows that the F1 component has a degree of circular polarisation of 25\% while the F2 component is up to 40\% circularly polarised which indicates that these two components are emitted at the fundamental plasma frequency (Figure 3). The H2 component is only up to 15\% circularly polarised in Figure 3, indicating that this component is emitted at the harmonic of the plasma frequency. All components are polarised in the same sense.F2 and H2 are therefore a fundamental-harmonic pair based on polarisation, frequency and imaging. There is, however, a delay in the fundamental F2 with respect to the harmonic H2 as seen in Figures 1b and 1c. This has been observed before in a similar study of a solar U-bursts \citep{do15}. \citet{do15} found that the delay appeared due to the difference in group velocities of the radio waves at the fundamental and the harmonic frequencies. }

{F1 and H1 can also constitute a fundamental-harmonic pair since F1 is highly polarised and H1 is present at twice the starting frequency of F1. Stokes V images from the NRH were used to estimate the degree of circular polarisation of the higher frequency components. At 150 and 173~MHz the radio sources are almost unpolarised which indicates harmonic plasma emission. This is consistent with the scenario that F1 and H1 are a fundamental-harmonic pair which is also supported by imaging (Figure 2b). However, at 228 and 298 MHz the radio sources are $\sim$15\% circularly polarised while the radio source at 270~MHz has a higher degree of circular polarisation of up to 50\%. The ORFEES dynamic spectrum (Figure 1b) shows a long-duration short-bandwidth burst around 228~MHz and noisy bursts at frequencies above. The high polarisation of the 270~MHz sources and the fact that these bursts occur at the start frequency of a J-burst could indicate the presence of radio spikes. }

\subsection{Jet characteristics and magnetic field configuration}

{The NRH sources at frequencies above 150~MHz are located directly on top of the solar jet (denoted by the white arrow in Figure 2). The jet was observed at EUV wavelengths in all AIA coronal filters. The evolution of the solar jet is shown in Figure 4 in AIA 171~\AA~ images as well as in the movie accompanying this paper. The jet eruption commenced at 11:06:23 UT only a few seconds after the onset of the Type III radio burst. The eruption lasted for $\sim$8~minutes which is a short duration compared to other jets studied \citep{mu16}. The material erupted appears to be directed southwards as seen in Figures 4c and 4d. We estimated the velocity of the jet to be 510~km~s$^{-1}$ in the horizontal direction, which is on the high end of the velocity scale of other jets observed \citep{mu16,yu14,ki07}. }

{The jet and the J-burst occur coincidently with a small B-class flare in the southern hemisphere as shown in Figure 5. The flare was associated with a soft X-ray source (6--12~keV) observed by the Reuven Ramaty High Energy Solar Spectroscopic Imager \citep[RHESSI;][]{li02} which was located in the southern hemisphere above the complex active region in Figure 2. The J-burst observed as the highest intensity peaks in the top panel of Figure 5 correlates very well in time with the rise phase of the jet (second panel) and does not seem to be related to the J-burst. The B-class flare can be seen in GOES 1--8~\AA~(third panel) and RHESSI 3--6 and 6--12~keV channels (bottom panel). The flare was brighter than the jet in the northern hemisphere at EUV wavelengths. No X-ray sources were associated with the jet in the northern hemisphere and therefore the jet was unrelated to this B-class flare. Radio imaging shows that the J-burst discussed here is associated with the jet in the northern hemisphere, as opposed to the B-class flare from the complex active region in the southern hemisphere. Despite the presence of a $\beta\gamma\delta$ active region in the south that has been producing numerous jets, this complex event, consisting of a jet and a J-burst, occurred in an area of much lower magnetic flux in the north.}

\begin{figure*}[ht]
\centering
\includegraphics[angle = 0, width = 370px, trim = 30px 0px 0px 30px ]{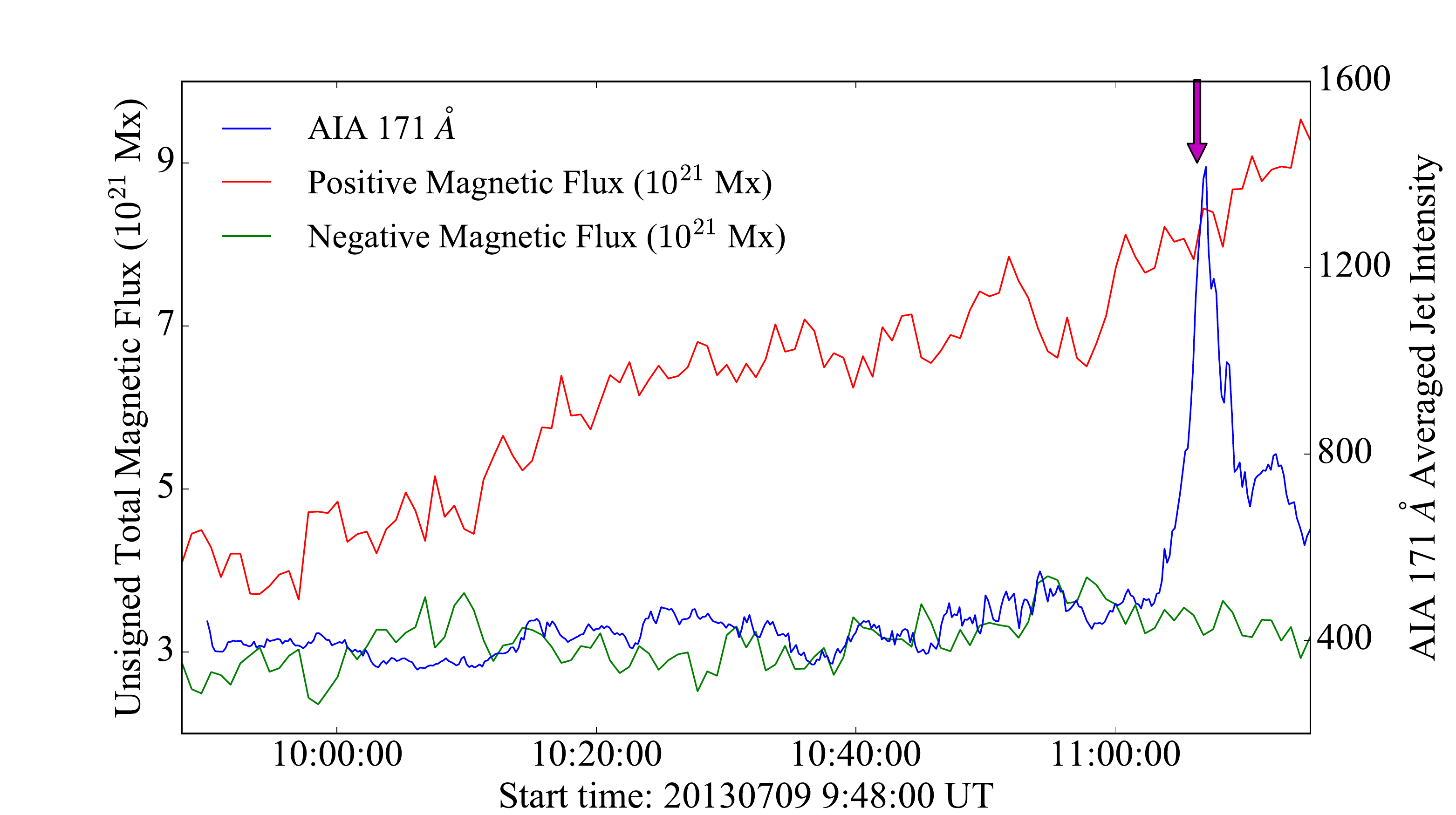}
\caption{Negative (green) and positive (red) magnetic flux estimated inside the dashed box in Figure 7 in which the positive polarity footpoint resides. The AIA 171 \AA~ light curve (blue) for the jet and accompanying bright loops is also plotted for comparison to show the onset and peak time of the jet activity. The purple arrow denotes the start of the J-burst followed by the jet moving southwards a few seconds after.  \label{fig5}}
\end{figure*} 

{The jet originates inside the negative polarity region of an extended area of bipolar plage in the northern hemisphere. This bipolar plage is connected via magnetic field lines to the group of complex $\beta\gamma\delta$ active regions in the southern hemisphere (Figure 6). PFSS models were used to estimate the connectivity between the area of bipolar plage and the active region. Figure 6 shows multiple magnetic field lines connecting these two regions, which are coincident with the locations of the LOFAR (blue) and NRH (red) radio source centroids at 150, 78 and 39~MHz, respectively from Figures 2b and 2c. Some of the magnetic field lines connecting southwards seen in Figure 6 could facilitate an escape medium for the J-burst producing electrons. The jet material also follows a southwards trajectory most likely due to magnetic field lines connecting southwards. } 

{A zoomed in view of the solar jet can be seen in Figure 7 where the positive (white) and negative (black) photospheric magnetic field contours are overlaid on top of AIA 171~\AA~images. The contours are plotted at the $\pm150$~G level. The panels of Figure 7 show the evolution of the region $\sim$1~hour before the jet occurred (10:00:00 UT), then $\sim$10 minutes before the jet (10:54:11 UT) and the third panel shows the occurrence of the solar jet (11:06:35 UT). }

{The jet originates next to a newly emerged positive polarity footpoint as seen in Figure 7. This positive polarity footpoint is also visible in the HMI inserts at the top left of each panel and it appears inside the negative polarity region of the plage. The footpoint does not appear as prominent one hour before the jet (Figure 7a), however it has already emerged 10 minutes before the jet when the plasma to the right of it brightens as seen in the AIA 171~\AA~ image (Figure 7b). The jet material is ejected to the right of the footpoint in the direction of the plasma brightenings but there may be projection effects. The jet then follows a southward trajectory along magnetic field lines in that direction. }

{Magnetic flux emergence, as seen inside the box in the panels of Figure 7, appears to be the reason the jet was triggered and electrons were accelerated to produce the J-burst. Figure 8 shows the positive (red) and negative polarity magnetic flux (green) that emerged from a small region inside the rectangular box in Figure 7 following the positive polarity footpoint. Also plotted in this figure is the integrated intensity of the solar jet in the AIA 171~\AA~filter (blue). The onset of the J-burst and the jet eruption are denoted by the purple arrow in Figure 8. We see a constant increase in the positive magnetic flux that keeps increasing even after the eruption of the jet. There seem to be no major changes in the negative flux. Over a time period of one hour, the positive magnetic flux becomes five times greater in this region. This observation is consistent with simulation results by \citet{sh92} where a solar jet was observed to occur as soon as 30 minutes after magnetic flux emergence. }

\section{Discussion and conclusion}

{The jet and J-burst in this analysis are related events caused by a sudden increase in the photospheric magnetic flux. These events are most likely a result of the magnetic reconnection between the newly emerged field lines and the overlying coronal magnetic field originating from the extended area of bipolar plage and not from an active region. The jet is associated with the emergence of positive polarity magnetic field inside a region of predominantly negative polarity. The jet occurs at the edge of the sunspot (Figure 7) which is similar to the jets observed by \citet{in11}. Since the surrounding area is predominantly of negative polarity, the newly emerged positive magnetic field lines have a high chance to reconnect with the overlying magnetic field lines. }

{Magnetic reconnection is therefore most likely to be responsible for the jet and accelerated electrons producing a bright J-burst with both fundamental and harmonic components. Components of this burst, such as F1, F2, H1 and, H2, were imaged by NRH and LOFAR over a broad frequency range of 35--173~MHz. The components F1 and H1 and F2 and H2 are fundamental-harmonic pairs based on their polarisation, co-spatial location and frequency range. However, F2 and H2 occur to the right of F1 and H1 and may be originating from a different electron beams escaping along different magnetic field lines. The co-spatial alignment of F1 and H1 and F2 and H2, respectively, as well as the polarisation of these components compares very well with previous studies by \citet{du80} where the fundamental-harmonic pairs studied were co-spatial and the fundamental was up to 35\% circularly polarised and the harmonic up to 11\% polarised. This study also confirms a delay in the fundamental with respect to the harmonic which has been observed before by \citet{do15}. The radio emission above 228~MHz that occured at the same spatial location as the jet, has some of the characteristics of radio spikes. The cut-off frequency of 33~MHz that was observed by both LOFAR and NDA can be explained if the burst travels along a closed magnetic loop until it reaches the top. }

{The distance to the top of the loop can be calculated using an electron density model that relates the electron density in the corona to distance from the photosphere. The electron density is related to the frequency of emission in the following way:
\begin{align}
\label{eq:1}
f = n f_p = n~C\sqrt{n_e} .
\end{align} 
The emission frequency, $f$, is given by the local plasma frequency, $f_p$, multiplied by the harmonic number $n$ and is directly proportional to the square root of the electron density, $n_e$, in cm$^{-3}$, where $C = 8980$~Hz~cm$^{3/2}$ is the constant of proportionality. We used the radial electron density model of \citet{sa77} to estimate the height at the emission at 33~MHz occurs (the frequency where F2 suddenly stops). We chose this density model as the J-burst does not to originate from an active region with high densities but from an area of bipolar plage with lower densities. The \citet{sa77} density model is suitable for estimates of the background electron densities in the solar corona and it will provided a lower limit of the height the electrons travelled when radio emission stopped. }

{The height corresponding to a frequency of 33~MHz is $\sim$360~Mm above the photosphere for fundamental emission which is equivalent to a distance of 1.52~R$_{\sun}$ from the solar centre. Electrons must travel along very long loops stretched out in the corona. Assuming the J-burst ends when the generating electrons reach the top of a magnetic loop or a region with a sudden change in magnetic field, a half loop length $>360$~Mm can be found in the numerous loops belonging to the trans-equatorial loop system that connects the area of bipolar plage in the northern hemisphere to the complex group of active regions in the southern hemisphere. This also agrees with the images of the bursts that place the emission sources above 150~MHz (Figure 6) closer to the jet and above the jet footpoint which corresponds to low heights in the corona. The emission below 80~MHz (Figure 6) appears to occur further away from the footpoint which corresponds to higher heights in the corona, however projection effects may be significant. There is a lot of connectivity between these two regions and a lot of magnetic loops at various heights can be found to explain the trajectory of the J-burst electrons. } 


{We, therefore, propose that the J-burst electrons were accelerated together with the solar jet, most likely by the magnetic reconnection between the overlying magnetic field lines and the newly emerged field lines. Radio spikes may have also accompanied this event due to the presence of highly polarised, higher frequency radio sources. The accelerated electrons travelled a large distance along a closed magnetic loop since radio emission stopped at a fixed frequency most likely at the top of this loop which is seen in LOFAR images and dynamic spectra. This study also shows non-radial trajectories of radio bursts as in \citet{mo14} and confirms some of the more interesting characteristics of a fundamental-harmonic pair such as the higher degree of polarisation \citep{du80} and the delay of the fundamental with respect to the harmonic \citep{do15}. It is interesting to note that such a complex small-scale event occurred away from the $\beta\gamma\delta$ active region present on the Sun but  in an area of relatively low magnetic flux. It is also, for the first time, that low frequency radio emission (<100~MHz) was spatially correlated with a solar jet and analysed in detail by combining NRH and LOFAR data. Type III bursts usually occur in the absence of flares 90\% of the time \citep{du85}, however such small-scale eruptive events arising due to magnetic reconnection could facilitate accelerated electrons continuously to produce the large numbers of Type III bursts observed in a similar way to the J-burst analysed here. }

{Radio emission was the best diagnostic in this analysis to trace accelerated electrons resulting from magnetic reconnection up to large heights in the corona. Broad frequency range observations that offer polarisation information (such as the combined LOFAR, NRH, ORFEES and NDA observations presented here) are needed in future observations to give a larger picture of flaring events associated with particle acceleration. Solar dedicated instruments, such as an European Solar Radio Array (ESRA), could facilitate these observations in the future.}

\begin{acknowledgements}{This work has been supported by a Government of Ireland studentship from the Irish Research Council (IRC), the Non-Foundation Scholarship awarded by Trinity College Dublin and the IRC New Foundations award. Bo Thid\'e was financially supported the Swedish Research Council (VR) under the contract number 2012-3297. The NRH is funded by the French Ministry of Education and the R\'egion Centre. ORFEES is part of the FEDOME project, partly funded by the French Ministry of Defense. LOFAR, designed and constructed by ASTRON, has facilities in several countries, that are owned by various parties (each with their own funding sources), and that are collectively operated by the International LOFAR Telescope (ILT) foundation under a joint scientific policy. We would like to acknowledge Valentin Melnik and Vladimir Dorovskyy for their comments and suggestions on the manuscript and supplying URAN-2 dynamic spectra for comparison. }\end{acknowledgements}

\bibliographystyle{aa} 
\bibliography{Paper4_AA.bib} 

\end{document}